# A General Systems Theory for the Observed Fractal Space-Time Fluctuations in Dynamical Systems

A. M. Selvam

Deputy Director (Retired)

Indian Institute of Tropical Meteorology, Pune 411 008, India

Email: amselvam@eth.net

Web sites: http://www.geocities.com/~amselvam

http://amselvam.tripod.com/index.html

## ABSTRACT

Recent studies of DNA sequence of letters A, C, G and T exhibit the inverse power law form frequency spectrum. Inverse power-law form of the power spectra of fractal space-time fluctuations is generic to the dynamical systems in nature and is identified as self-organized criticality. In this study it is shown that the power spectra of the frequency distributions of bases C+G in the Human chromosome X DNA exhibit self-organized criticality. DNA is a quasicrystal possessing maximum packing efficiency in a hierarchy of spirals or loops. Self-organized criticality implies that non-coding *introns* may not be redundant, but serve to organize the effective functioning of the coding *exons* in the DNA molecule as a complete unit.

## 1. INTRODUCTION

Animate and inanimate structures in nature exhibit selfsimilarity in geometrical shape [1], i.e. parts resemble the whole object in shape. The most fundamental selfsimilar structure is the forking (bifurcating) structure of tree branches, lung architecture, river tributaries, branched lightning, etc. The hierarchies of complex branching structures represent uniform stretching on a logarithmic scale and belong to the recently identified [2] category of fractal objects. The complex branching architecture is a selfsimilar fractal since branching occurs on all scales (sizes) and forms the geometrical shape of the whole object. The substantial identity of the fractal architecture underlying leaf and branch was recognized more than three centuries ago [3]. This type of scaling was used by D'Arcy Thompson [4] in scaling anatomical structures. It appears quite often in the form of allometric growth laws in botany as well as in biology [5]. This particular kind of scaling has been successfully used in biology for over a century. The universal symmetry of selfsimilarity underlies apparently irregular complex structures in nature [6]. The study of fractals now belongs to the new field *'nonlinear dynamics and chaos'*, a multidisciplinary area of intensive research in all fields of science in recent years (since 1980s) [7].

Selfsimilar growth, ubiquitous to nature is therefore governed by universal dynamical laws, which are independent of the exact details (chemical, physical, physiological, electrical etc.) of the dynamical system. Dynamical systems in nature possess selfsimilar fractal geometry to the spatial pattern which support fluctuations of processes at all time scales. The temporal fluctuations of dynamical processes are also scale invariant in time and are therefore temporal fractals. The fractal structure to the spatial pattern was pursued as an independent multidisciplinary area of research since late 1970s when the concept of fractals was introduced by Mandelbrot [2]. Though the scale invariant (selfsimilar) characteristic of temporal fluctuations of dynamical systems has a longer history of observation of more than 30 years [8], it was not studied in relation to the associated fractal architecture of spatial pattern till as late as 1988. The unified concept of space-time fractals was identified as a signature of self-organized criticality by Bak *et al* [9]. For example, the fluctuations in time of atmospheric flows, as recorded by meteorological parameters such as pressure, temperature, wind speed etc., exhibit selfsimilar fluctuations in time namely, a zigzag pattern of increase (decrease) followed by a decrease (increase) on time scales from seconds to years. Such jagged pattern for atmospheric variability (temporal) resembles the selfsimilar coastline (spatial) structure. Selfsimilarity indicates long-range correlations, i.e. the amplitude of short- and long-term fluctuations are related to each other by a non-dimensional scale factor alone. Therefore, dynamical laws which govern the space-time fluctuations of smallest scale (turbulence, millimeters-seconds) fluctuations in space-time also apply for the largest scale (planetary, thousands of kilometers - years) in atmospheric flows throughout the globe.

Biological systems exhibit high degree of co-operation in the form of long-range communication. The concept of co-operative existence of fluctuations in the organization of coherent structures has been identified as selforganized co-operative phenomena [10] and synergetics [11]. The physics of self-organized criticality exhibited by dynamical systems in nature is not yet identified. Selvam [12] has developed a cell dynamical system model for atmospheric flows which shows that the observed long-range spatiotemporal correlations namely, self-organized criticality are intrinsic to quantumlike mechanics governing turbulent flow dynamics. The model concepts are independent of the exact details, such as the chemical, physical, physiological, etc. properties of the dynamical systems and therefore provide a general systems theory [13,14,1,15] applicable for all dynamical systems in nature.

## 2. MODEL CONCEPTS

Power spectra of fractal space-time fluctuations of dynamical systems such as fluid flows, stock market price fluctuations, heart beat patterns, etc., exhibit inverse power-law form identified as self-organized criticality [9] and represent a selfsimilar eddy continuum. Li (2004)[16] has given an extensive and informative bibliography of the observed $1/f$ noise, where $f$ is the frequency, in biological, physical, chemical and other dynamical systems. A general systems theory [12,17,18] developed by the author shows that such an eddy continuum can be visualised as a hierarchy of successively larger scale eddies enclosing smaller scale eddies. Since the large eddy is the integrated mean of the enclosed smaller eddies, the eddy energy (variance) spectrum follows the statistical normal distribution according to the Central Limit Theorem [19]. Hence the additive amplitudes of eddies, when squared, represent the probabilities, which is also an observed feature of the subatomic dynamics of quantum systems such as the electron or photon [20,21]. The long-range correlations intrinsic to self-organized criticality in dynamical systems are signatures of quantumlike chaos associated with the following characteristics: (a) The fractal fluctuations result from an overall logarithmic spiral trajectory with the quasiperiodic Penrose tiling pattern [12,17,18] for the enclosed structure. (b) Conventional continuous periodogram power spectral analyses of such spiral trajectories will reveal a continuum of wavelengths with progressive increase in phase. (c) The broadband power spectrum will have embedded dominant wavebands, the bandwidth increasing with wavelength, and the wavelengths being functions of the golden mean. The first 13 values of the model predicted [12,17,18] dominant peak wavelengths are 2.2, 3.6, 5.8, 9.5, 15.3, 24.8, 40.1, 64.9, 105.0, 167.0, 275, 445.0 and 720 in units of the block length 10bp (base pairs) in the present study. Wavelengths (or periodicities) close to the model predicted values have been reported in weather and climate variability [15], prime number distribution [22], Riemann zeta zeros (non-trivial) distribution [23], stock market economics [24,25], DNA base sequence structure [26]. (d) The conventional power spectrum plotted as the variance versus the frequency in log-log scale will now represent the eddy probability density on logarithmic scale versus the standard deviation of the eddy fluctuations on linear scale since the logarithm of the eddy wavelength represents the standard deviation, i.e. the r.m.s (root mean square) value of the eddy fluctuations. The r.m.s. value of the eddy fluctuations can be represented in terms of statistical normal distribution as follows. A normalized standard deviation $t=0$ corresponds to cumulative percentage probability density equal to 50 for the mean value of the distribution. For the overall logarithmic spiral circulation the logarithm of the wavelength represents the r.m.s. value of eddy fluctuations and the normalized standard deviation $t$ is defined for the eddy energy as

$$ t = \frac{\log_e L}{\log_e T_{50}} - 1 \qquad (1) $$

The parameter $L$ in Eq. 1 is the wavelength and $T_{50}$ is the wavelength up to which the cumulative percentage contribution to total variance is equal to 50 and $t = 0$. The variable $\log_e T_{50}$ also represents the mean value for the r.m.s. eddy fluctuations and is consistent with the concept of the mean level represented by r.m.s. eddy fluctuations. Spectra of time series of fluctuations of dynamical systems, for example, meteorological parameters, when plotted as cumulative percentage contribution to total variance versus $t$ follow the model predicted universal spectrum [17; see all references under Selvam].

## 3. DATA AND ANALYSIS

The Human chromosome X DNA base sequence was obtained from the entrez Databases, Homo sapiens Genome (build 34, version 1) at http://www.ncbi.nlm.nih.gov/entrez. The first 4 contiguous data sets (Table 1) were chosen for the study. The number of times base C and also base G, i.e., (C+G), occur in successive blocks of 10 bases were determined in successive length sections of 70000 base pairs giving a series of 7000 values for each data set giving respectively 10, 8 and 19 data sets for the contiguous data sets 2 to 4. A data series of 8652 C+G concentration per successive 10 bp values was used for the short contiguous data set 1.

The power spectra of frequency distribution of bases were computed accurately by an elementary, but very powerful method of analysis developed by Jenkinson [27] which provides a quasi-continuous form of the classical periodogram allowing systematic allocation of the total variance and degrees of freedom of the data series to logarithmically spaced elements of the frequency range (0.5, 0). The cumulative percentage contribution to total variance was computed starting from the high frequency side of the spectrum. The power spectra were plotted as cumulative percentage contribution to total variance versus the normalized standard deviation $t$. The average variance spectra for the Human chromosome X DNA data sets (Table 1) and the statistical normal distribution are shown in Fig. 1. The 'goodness of fit' (statistical chi-square test) between the variance spectra and statistical normal distribution is significant at less than or equal to 5% level for all the data sets (Table 1).

The power spectra exhibit dominant wavebands where the normalized variance is equal to or greater than 1. The dominant peak wavelengths were grouped into 13 class intervals $2 - 3$, $3 - 4$, $4 - 6$, $6 - 12$, $12 - 20$, $20 - 30$, $30 - 50$, $50 - 80$, $80 - 120$, $120 - 200$, $200 - 300$, $300 - 600$, $600 - 1000$ (in units of 10bp block lengths) to include the model predicted dominant peak length scales mentioned earlier. Average class interval-wise percentage frequencies of occurrence of dominant wavelengths are shown in Fig. 2 along with the percentage contribution to total variance in each class interval corresponding to the normalised standard deviation $t$ computed from the average $T_{50}$ (Fig. 2) for each data set. The average and standard deviation of the wavelength $T_{50}$ up to which the cumulative percentage contribution to total variance is equal to 50 are also shown in Fig. 2.

## 4. RESULTS AND CONCLUSIONS

The limited number of only four contiguous data sets used in the present study give the following results. The variance spectra for almost all the data sets exhibit the universal inverse power-law form $1/f^\alpha$ of the statistical normal distribution (Fig. 1) where $f$ is the frequency and the spectral slope $\alpha$ decreases with increase in wavelength and approaches 1 for long wavelengths. The above result is also seen in Fig. 2 where the wavelength class interval-wise percentage frequency distribution of dominant wavelengths follow closely the corresponding computed variation of percentage contribution to the total variance as given by the statistical normal distribution. Inverse power-law form for power spectra implies long-range spatial correlations in the frequency distributions of the bases C+G in DNA. Li and Holste [28] have recently identified universal $1/f$ spectra and diverse correlation structures in Guanine (G) and Cytosine (C) content of human chromosomes. Microscopic-scale quantum systems such as the electron or photon exhibit non-local connections or long-range correlations and are visualized to result from the superimposition of a continuum of eddies. Therefore, by analogy, the observed fractal fluctuations of the frequency distributions of the bases exhibit quantumlike chaos in the Human chromosome X DNA. The eddy continuum acts as a robust unified whole fuzzy logic network with global response to local perturbations. Therefore, artificial modification of base sequence structure at any location may have significant noticeable effects on the function of the DNA molecule as a whole. Results of the limited number of data sets in this study indicates that the presence of introns, which do not have meaningful code, may not be redundant, but may serve to organize the effective functioning of the coding exons in the DNA molecule as a complete unit.

## 5. ACKNOWLEDGEMENT

The author is grateful to Dr. A. S. R. Murty for encouragement.

Human Chromosomes", Submitted for Journal publication, 2004.

**variance and phase spectra - human chromosome X DNA bases C+G frequency distribution**

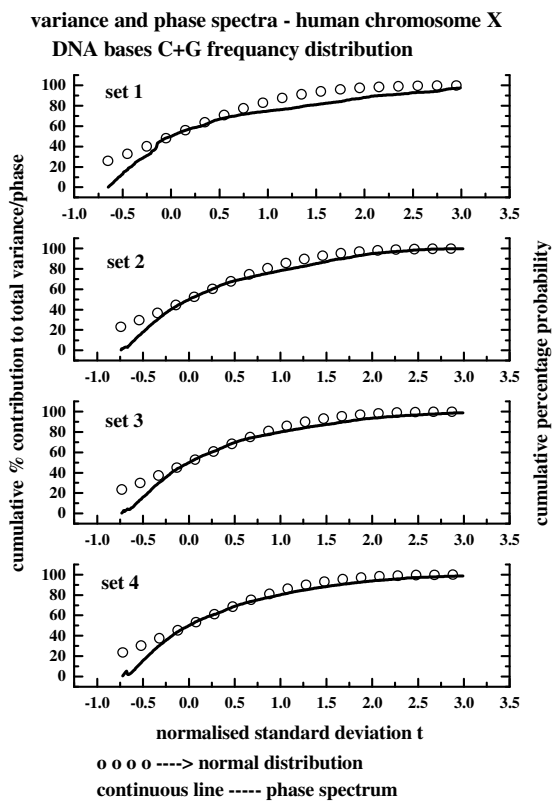

o o o o ----> normal distribution

continuous line ----- phase spectrum

Figure 1: Average variance spectra for the concentration of base combination C+G per 10 successive base pairs in Human chromosome X DNA. Continuous lines represent the variance spectra and open circles give the statistical normal distribution.

**class interval-wise percentage frequency of occurrence of dominant eddies**

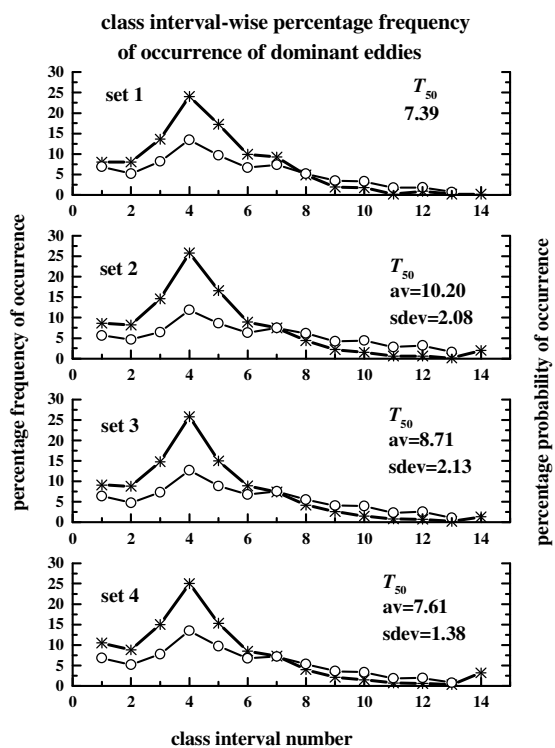

Figure 2: Average wavelength class interval-wise percentage distribution of dominant (normalized variance greater than 1) wavelengths are given by line + star. The computed percentage contribution to the total variance for each class interval is given by line + open circle. The mean and standard deviation of the wavelengths $t_{50}$ up to which the cumulative percentage contribution to total variance is equal to 50 are also given in the figure.

TABLE 1

| s.no | Accession number | Base pairs | | Base pairs used for analysis | | Number of data sets | Mean C+G concentration per 10bp | Spectra following normal distribution (%) |
|---|---|---|---|---|---|---|---|---|
| | | from | to | from | to | | | |
| 1 | NT_078115.2 | 1 | 86563 | 1 | 86520 | 1 | 5.46 | 100 |
| 2 | NT_028413.7 | 1 | 766173 | 1 | 700000 | 10 | 4.79 | 100 |
| 3 | NT_033330.6 | 1 | 623707 | 1 | 560000 | 8 | 4.87 | 100 |
| 4 | NT_025302.12 | 1 | 1381418 | 1 | 1330000 | 19 | 4.38 | 100 |